\documentclass[aps,pra,twocolumn,superscriptaddress]{revtex4-1}

\usepackage{graphicx}
\usepackage{dcolumn}
\usepackage{bm}
\usepackage{amsmath}
\usepackage{amssymb}
\usepackage{latexsym}
\usepackage{epsfig}
\usepackage{amsbsy}
\usepackage{array}
\usepackage{amssymb}
\usepackage{setspace}
\usepackage{bm}

\begin{document}

\title{The defect-induced localization in many positions of the quantum random walk }

\author{Tian Chen$^{}$}
\email{chentian@bit.edu.cn}
\author{Xiangdong Zhang$^{}$}

\affiliation{$^{}$School of Physics, Beijing Institute of Technology, Beijing, 100081, China\\
}

\begin{abstract}
We study the localization of probability distribution in a discrete quantum random walk on an infinite chain. With a phase defect introduced in any position of the quantum random walk (QRW), we have found that the localization of the probability distribution in the QRW emerges. Different localized behaviors of the probability distribution in the QRW are presented when the defect occupies different positions. Given that the coefficients of the localized stationary eigenstates relies on the coin operator, we reveal that when the defect occupies different positions, the amplitude of localized probability distribution in the QRW exhibits a non-trivial dependence on the coin operator.

\end{abstract}


\maketitle
\section{Introduction}
The classical random walk (CRW) has proven to be a powerful technique in classical algorithms~\cite{Motwani_book95}. Its quantum counterpart, quantum random walk (QRW)~\cite{Kempe_CP2003, Andraca_QIP2012, Farhi_PRA1998, Aharonov_PRA1993, Childs_QIP2002, Strauch_PRA2006}, has also been employed in designing quantum algorithms and displayed its advantage in the universal quantum computation~\cite{Shenvi_PRA2003, Keating_PRA2007, Perets_PRL2008,Childs_PRL2009, Zahringer_PRL2010, Lovett_PRA2010, Underwood_PRA2010, Yalcinkaya_JPA2015}. Another interesting aspect of study in the quantum walk architecture is the localization of position distribution~\cite{Wojcik_PRL2004, Ribeiro_PRL2004, Inui_PRA2004}. Such study of localization has its applications in the exploration of properties of low dimensional materials, i.e., Anderson localization~\cite{Anderson_PR1958}, and the understanding of the building quantum walk algorithms and quantum computations~\cite{Keating_PRA2007}, etc. Many different methods have been proposed to realize the localization of position distribution in the QRW architecture. With the introduction of static or dynamic disorder into the standard QRW, the interference patten of the QRW becomes changed, the position distribution in the walk shows diffusive spreading and the localization effect under some conditions, not the ballistic spreading in the standard QRW~\cite{Wojcik_PRL2004, Ribeiro_PRL2004, Inui_PRA2004, Konno_QIP2010, Chandrashekar_PRA2011, Crespi_Nat2013, Xue_NJP2014, Nicola_PRA2014, Xue_PRL2015}. Moreover, by modifying the phase of the positions of the QRW, we can localize the particle in some certain positions, and the probability distribution for the different positions of the QRW displays a periodical behavior with time~\cite{Wojcik_PRL2004, Xue_NJP2014}. Recent researches illustrated that when only the phase of the original position of the QRW is modified (it means that only one single phase defect is introduced at the original position), one will obtain a sharp allocation of distribution for this particular position in our QRW at the infinite time~\cite{Wojcik_PRA2012, Li_PRA2013, Zhang_PRA2014, Lam_PRA2015, Zhang_QIP2014, Li_SRep2015}. This QRW incorporating one position-dependent phase defect has been realized with the aid of beam displacers and phase shifters in experiment already~\cite{Xue_SRep2014, Xue_PRA2015}.

In our work, we study the localization of the position distribution of a QRW on an infinite line. The inhomogeneity is introduced with one phase defect residing at the position $x=n$.
As stated in the previous paper, if the defect occupies the position $x=0$ or $x=1$, the probability distribution for certain positions in the QRW architecture will not tend to zero even the time approaches the infinite limit~\cite{Wojcik_PRA2012, Zhang_PRA2014, Xue_SRep2014}. Our results reveal that, when the defect is introduced into any position of the QRW, the localization of the probability distribution will appear with our chosen initial state. The amplitude of localized probability relies on the overlap between the localized stationary eigenstates and the initial state.
We find that, at the position where the defect occupies, the amplitude of the probability distribution of this QRW exhibits the non-monotonic increase with the increase of the parameter $\theta$ of the coin operator $C(\theta)$, while in comparison, the probability of occupying the defect's position at $x=0$ or $x=1$ shows a linear increase with the parameter $\theta$~\cite{Xue_SRep2014}. Given that the expressions for localized stationary eigenstates of the QRW with different coin operators have been obtained in our work, we present a reasonable analysis for this non-trivial dependence on the coin operator. The potential experimental realization of our QRW with the phase defect is proposed at the end.

The organization of our work is as follows, Sec. \ref{II} introduces our QRW architecture with one single phase defect. The localized stationary states with different coin operators are presented. In Sec. \ref{III}, we study the position distribution of the QRW with the defect. We take the defect residing at the position $x=2$ or $x=3$ as examples, and analyze the effect of the coin operator on the position distribution in the QRW. Then we propose a potential experimental realization for the QRW. We conclude in Sec. \ref{IV}.

\section{the quantun random walk with different coin operators}\label{II}
The one step in the QRW architecture is represented as $U_\phi$ which consists of one coin operator $C(\theta)$ and one conditional shift operator $S_c^{\phi}$.
\begin{equation}
U_{\phi}=(\sum_{c=0,1}|c\rangle\langle c|\bigotimes S_c^{\phi})(C(\theta)\bigotimes\mathcal{I}),
\end{equation}
where the Hilbert space of coin system $\mathcal{H}_c$ is spanned by $|c\rangle$, $c=0,1$, and the Hilbert space of position $\mathcal{H}_p$ is spanned by $|x\rangle$, $x\in \mathbf{Z}$. The total system is comprised by the coin and the position. The coin operator $C(\theta)$ is $\theta$-dependent, that is,
\begin{equation}
H=C(\theta)=\left(\begin{array}{cc}
\cos\theta & \sin\theta\\
\sin\theta & -\cos\theta\end{array}\right).
\end{equation}
When $\theta=\pi/4$, the coin operator takes the form as the familiar Hadamard matrix. The conditional shift operator $S_c^{\phi}$ allows the particle to walk into different directions according to the coin state,
\begin{equation}
\begin{split}
S_{0}^{\phi}|m\rangle=e^{2\pi i\phi\delta_{x,m}}|m-1\rangle,\\
S_{1}^{\phi}|m\rangle=e^{2\pi i\phi\delta_{x,m}}|m+1\rangle.
\end{split}
\end{equation}
The effect of the defect is contained in the phase. When the particle walks through the position $x=m$, it will acquire an additional phase $2\pi i\phi$. We assume that the state of the position and the coin is,
\begin{equation}
|\psi\rangle=\sum_n(\alpha_n|0\rangle_c|n\rangle_p+\beta_n|1\rangle_c|n\rangle_p).\label{state}
\end{equation}
The subscript $c$ $(p)$ indicates that this state belongs to the Hilbert space for the coin (position). After applying one step $U_\phi$ to the total system, we obtain the expressions of the amplitude $\alpha_n$ and $\beta_n$ when the particle starts from the position $n=m$ at the discrete time $t$,
\begin{subequations}
\begin{align}
\alpha_{m-1}(t+1)=\omega\cdot\alpha_m(t)\cdot\cos\theta+\omega\cdot\beta_m(t)\cdot\sin\theta,\\
\beta_{m+1}(t+1)=\omega\cdot\alpha_m(t)\cdot\sin\theta-\omega\cdot\beta_m(t)\cdot\cos\theta;
\end{align}
\end{subequations}
Here, the parameter $\omega$ denotes the phase $e^{2\pi i\phi}$, with $\phi\in[0,1)$. When the particle starts from the position, $n\neq m$ at time $t$, the time evolution for coefficients $\alpha_n$ and $\beta_n$ are,
\begin{subequations}
\begin{align}
\alpha_{n}(t+1)=\alpha_{n+1}(t)\cdot\cos\theta+\beta_{n+1}(t)\cdot\sin\theta,\\
\beta_{n}(t+1)=\alpha_{n-1}(t)\cdot\sin\theta-\beta_{n-1}(t)\cdot\cos\theta.
\end{align}
\end{subequations}
Considering the particle starts from the original position ($x=0$) initially, it is clearly that the particle occupies even (odd) positions of the QRW architecture when the particle takes the even (odd) steps. To find the localized stationary states of the QRW with the defect, we apply two-step evolution operator $U_\phi^2$ for the total system. Following the method presented in Ref.~\cite{Wojcik_PRA2012}, the probability amplitude $\alpha_n$ and $\beta_n$ ($n\neq m$) can be obtained as,
\begin{subequations}\small
\begin{align}
\alpha_n(t+2)=&\alpha_{n+2}(t)\cdot\cos^2\theta+\beta_{n+2}(t)\sin\theta\cdot\cos\theta\notag\\&+\alpha_n(t)\sin^2\theta-\beta_n(t)\sin\theta\cdot\cos\theta=\lambda\cdot\alpha_n(t),\label{dyn_alpha}\\
\beta_n(t+2)=&\alpha_n(t)\cdot\cos\theta\cdot\sin\theta+\beta_n(t)\sin^2\theta\notag\\&-\alpha_{n-2}(t)\cdot\sin\theta\cdot\cos\theta+\beta_{n-2}(t)\cdot\cos^2\theta=\lambda\cdot\beta_n(t).\label{dyn_beta}
\end{align}
\end{subequations}
Here, the parameter $\lambda$ stands for the eigenvalues of $U_\phi^2$. Combining Eq.~\ref{dyn_alpha} and Eq.~\ref{dyn_beta}, we can get,
\begin{equation}
\beta_{n+2}=\frac{\alpha_{n+2}-\lambda\cdot\alpha_n}{\lambda-1}\cot\theta.\label{beta}
\end{equation}
Substituting the expression of $\beta_{n}$ into Eq.~\ref{dyn_beta}, we achieve the expression as,
\begin{equation}
\begin{split}
\lambda\cos^2\theta\alpha_{n+2}-(\lambda^2-2\lambda\sin^2\theta+1)\alpha_n+\lambda&\cos^2\theta\alpha_{n-2}=0,\\
&\quad n\neq m\pm2, m.\label{evol}
\end{split}
\end{equation}
The general solution of this equation is,
\begin{equation}
\alpha_n=C_+\cdot z^{n-m}+C_-\cdot z^{-(n-m)},\label{solu}
\end{equation}
where $C_+$ and $C_-$ are constant coefficients. Considering the convergence of $\alpha_n$ when $n\rightarrow\pm\infty$, we can obtain the expression for $\alpha_n$ with substituting Eq.~\ref{solu} into Eq.~\ref{evol},
\begin{subequations}
\begin{align}
&\alpha_n=C_+\cdot z^{n-m},\quad n\geq m+2,\\
&\alpha_n=C_-\cdot z^{-(n-m)},\quad n\leq m-2.
\end{align}
\end{subequations}
Here, $z$ is the solution of Eq.~\ref{evol} when its value satisfies, $|z|<1$. With replacing the expressions of $\alpha_n$ above into into Eq.~\ref{beta}, we can obtain $\beta_n$ as,
\begin{subequations}
\begin{align}
&\beta_{n+2}=C_+\frac{z^2-\lambda}{\lambda-1}z^{n-m}\cdot\cot\theta,\quad n\geq m+2,\\
&\beta_{n+2}=C_-\frac{1-\lambda\cdot z^2}{\lambda-1}z^{-(n-m+2)}\cdot\cot\theta,\quad n\leq m-4.
\end{align}
\end{subequations}
Taking into account the additional phase acquired when the particle walks through the defect at position $x=m$, we can get the coupled equations for the probability amplitude $\alpha_n$ and $\beta_n$ ($n=m$) with the evolution operator $U_\phi^2$ as,
\begin{subequations}\small
\begin{align}
\alpha_{m+2}\cos^2\theta+\beta_{m+2}\sin\theta\cos\theta+\omega\alpha_{m}\sin^2\theta-\omega\beta_m&\sin\theta\cos\theta\notag\\=&\lambda\alpha_m,\label{zero_1}\\
\omega\alpha_m\cos\theta\sin\theta+\omega\beta_m\sin^2\theta-\alpha_{m-2}\sin\theta\cos\theta+&\beta_{m-2}\cos^2\theta\notag\\=&\lambda\beta_m,\label{zero_2}\\
\omega\alpha_{m}\cos^2\theta+\omega\beta_{m}\sin\theta\cos\theta+\alpha_{m-2}\sin^2\theta-\beta_{m-2}&\sin\theta\cos\theta\notag\\=&\lambda\alpha_{m-2},\\
\alpha_m\cos\theta\sin\theta+\beta_m\sin^2\theta-\omega\alpha_{m}\sin\theta\cos\theta+\omega\beta_m&\cos^2\theta\notag\\=&\lambda\beta_{m+2}.
\end{align}
\end{subequations}
The parameter $\lambda$ stands for the eigenvalues of $U_\phi^2$, also. Following the obtained equations above, the explicit expressions for $\alpha_m$ and $\beta_m$ are,
\begin{subequations}
\begin{align}
&\alpha_m=\frac{\alpha_{m+2}}{\lambda}+\frac{1-\lambda}{\lambda}\beta_{m+2}\cdot\tan\theta,\\
&\beta_m=\frac{\lambda-1}{\lambda}\tan\theta\cdot\alpha_{m-2}+\frac{1}{\lambda}\beta_{m-2}.
\end{align}
\end{subequations}
With the aid of representations as, $\alpha_{m+2}=C_+\cdot z^2$, $\alpha_{m-2}=C_-\cdot z^2$, $\beta_{m+2}=C_+\frac{z^2-\lambda}{\lambda-1}\cot\theta$, and $\beta_{m-2}=C_-\frac{1-\lambda z^2}{\lambda-1}z^2\cdot\cot\theta$, we obtain the probability distribution $\alpha_m$ and $\beta_m$ at the position $x=m$ that the defect occupies as,
\begin{subequations}
\begin{align}
&\alpha_m=C_+,\\
&\beta_m=C_-\frac{1-\lambda z^2}{\lambda-1}\cot\theta.
\end{align}
\end{subequations}
We replace the term $\alpha_n$ and $\beta_n$ of Eq.~\ref{zero_1} and Eq.~\ref{zero_2} with the expressions above. The relation between the phase $\omega$ induced by the defect, the eigenvalue $\lambda$ of $U_\phi^2$ and the angle $\theta$ of the coin operator is shown as,
\begin{equation}\small
\begin{split}
\omega^2\cdot\sin^2\theta\cdot\cos^2\theta&=(y\cdot\cos^2\theta+\frac{y-\lambda}{\lambda-1}\cos^2\theta+\omega\cdot\sin^2\theta-\lambda)\\&(\lambda-y\cdot\cos^2\theta-\omega\cdot\sin^2\theta+
\frac{\lambda-1}{1-\lambda y}y\cdot\sin^2\theta),\label{lamb}
\end{split}
\end{equation}
where we use $y$ to replace $z^2$. The expression for $y$ can be obtained as,
\begin{equation}
y=\frac{\lambda^2+\omega^2-2\lambda\cdot\omega\cdot\sin^2\theta}{\lambda+\lambda\cdot\omega^2-2\lambda\cdot\omega\cdot\sin^2\theta}.
\end{equation}
The relation between the constants $C_+$ and $C_-$ is,
\begin{equation}\small
\begin{split}
&C_-\\=&C_+\frac{\cos^2\theta\cdot(\lambda-\omega^2)+(\omega\cdot\sin^2\theta-\lambda)(1+\omega^2-2\omega\cdot\sin^2\theta)}{\omega\cdot\cos^2\theta\cdot(2\omega\cdot\sin^2\theta-\lambda-1)}.
\end{split}
\end{equation}
Considering the normalized condition for the summation of $|\alpha_n|^2$ and $|\beta_n|^2$, we can get the values of $C_+$ and $C_-$, and the coefficients $\alpha_n$ and $\beta_n$ at different positions $n$ can be obtained.

\section{Results and Discussion}\label{III}
\subsection{Localization with the defect occupying different positions}
\begin{figure}[htbp]
\begin{center}
\includegraphics[width=0.5\textwidth]{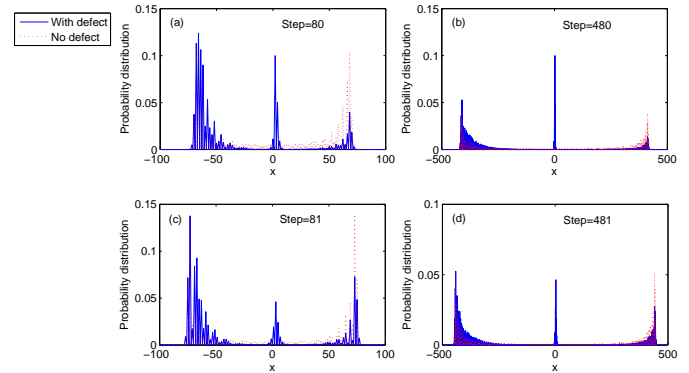}
\end{center}
\caption{\label{fig1} (Color online): The probability distribution of the position in the QRW with different steps. Blue solid, the probability distribution of the QRW with one single phase defect. Red dotted, the probability distribution of the standard QRW without defects. The initial state of the coin and position is taken as, $|\Phi\rangle_{ini}=(\frac{i}{\sqrt{2}}|0\rangle+\frac{1}{\sqrt{2}}|1\rangle)_{c}|0\rangle_p$. The phase of the defect, $\phi=1/2$. Figure (a) and (b), the defect occupies the position $x=2$. A sharp peak of probability is found at position $x=2$. The parameter $\theta$ of the coin operator is chosen, $\theta=\pi/6$. Figure (c) and (d), the defect occupies the position $x=3$. A sharp peak of probability is found at position $x=3$. The parameter $\theta$ of the coin operator is chosen, $\theta=\pi/8$.}
\end{figure}

In the following, we study the localization of probability distribution of the QRW at different positions. As well known that, if the particle starts from the original position ($x=0$), the particle will occupy only even (odd) positions with the even (odd) step evolution. In our numerical calculation, same probability distributions of the QRW with or without defects can be obtained when the defect occupies the position $x=0$ or $x=1$~\cite{Wojcik_PRA2012, Li_PRA2013, Zhang_PRA2014}. We explore the properties of probability distribution of the QRW in which the defect occupies a farther position ($x\geq2$) next. We take the single phase defect locating at the position $x=2$ or $x=3$ as examples. The particle starts from the original point, $x=0$, then undergoes many steps of the evolution in the quantum walk architecture with the single phase defect occupying the position $x=2$ or $x=3$. The probability distributions of the QRW with and without defects are presented in Fig.~\ref{fig1}.
It is clearly seen that when the defect appears, the probability of occupying the position around $x=2$ (Fig.~\ref{fig1} (a), (b)) or $x=3$ (Fig.~\ref{fig1} (c), (d)) does not tend to zero, no matter how many steps the particle has taken (Blue solid lines in Fig.~\ref{fig1}). The probability locating at the position $x=2$ or $x=3$ keeps the same value with the increase of steps. While, the probability distribution of the standard QRW without defects shows the ballistic spreading. No localization of probability distribution in the QRW can be found in such case (Red dotted lines in Fig.~\ref{fig1}). As stated previously~\cite{Wojcik_PRA2012}, the localization of the probability distribution in the QRW means that the amplitude of probability at certain position will not tend to zero with time changing. Due to the emergence of localized eigenstates induced in the QRW with defects~(see Sec.~\ref{II} above of our paper and related parts of Ref.~\cite{Wojcik_PRA2012, Zhang_PRA2014}), the localization in the QRW appears with the appropriate choice of initial states. Though in small steps of evolution, the localized probability in the QRW where the defect occupies the position $x=2$ or $x=3$ mingles with the diffusion of the probability, the localization becomes apparent when the steps of evolution is large. Another interesting feature in the QRW with defects is needed to be mentioned that the probability distribution of the QRW exhibits an asymmetrical distribution around the defect's position $x=2$ (from Fig.~\ref{fig1} (a) to (b)). Due to the reflection of the defect, a larger probability distribution can be found in the left side of the position $x=2$, when compared to the smaller probability of transmission in the right region of the position $x=2$~\cite{Li_PRA2013}. The similar behaviors of probability distribution can be found when the defect occupies the position $x=3$, see Fig.~\ref{fig1} (c) and (d).

Next, we discuss the amplitude of localized probability at the position where the defect occupies. Different positions are chosen. The particle starts from the original position ($x=0$).
\begin{figure}[htbp]
\begin{center}
\includegraphics[width=0.5\textwidth]{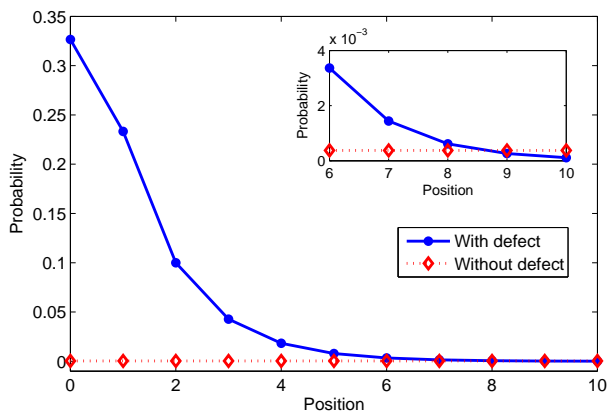}
\end{center}
\caption{\label{fig2} (Color online): The amplitude of localized probability at the position where the defect occupies. For the defect resides at the even positions, the step of the evolution in the QRW is 980; for the defect occupies the odd position, the step of the evolution in the QRW is 981. The initial state of the coin and position is taken as, $|\Phi\rangle_{ini}=(\frac{i}{\sqrt{2}}|0\rangle+\frac{1}{\sqrt{2}}|1\rangle)_{c}|0\rangle_p$. The phase of the defect, $\phi=1/2$. The parameter $\theta$ of the coin operator is chosen, $\theta=\pi/6$. Blue dotted, the QRW with one defect; red diamond, the QRW without defects. The figure inset reveals the amplitudes of localized probability for the position of the defect changes from 6 to 10.}
\end{figure}
From Fig.~\ref{fig2}, when the evolution of QRW is executed about 1000 steps, we find that the localization of probability distribution emerges at the certain position where the defect occupies. Though the magnitude of probability localized at the position $x\geq6$ is small, such probability will never decrease to zero with the increasing the step evolution of the QRW. As presented in inset of Fig.~\ref{fig2}, when the step of the evolution is around 1000, the probability at the position $x=9$ or $x=10$ of the QRW with defects is smaller than that of standard QRW without defects. With the increase of step of the evolution, the localized probability at the position $x=9$ or $x=10$ of the QRW with defects keep the same value, while the amplitude of probability for the standard QRW at position $x=9$ or $x=10$ decay to zero. It indicates that the localization of probability distribution in the QRW with defects appears indeed.

\subsection{The effect of coin operators on the localization}
We study the effect of different coin operators on the localization in the QRW. In our discussion, the one step of evolution $U_\phi$ contains the $\theta$-dependent coin operator ($C(\theta)$), followed by the conditional shift operator. To analyze the localization of probability distribution with the localized eigenstate of the QRW, we need to consider the operator $U_\phi^2$ containing the two-step evolution in the QRW. We explore the properties of the probability distribution in the QRW with the even step evolution at first, then the odd step evolution of the QRW is discussed.

\begin{figure}[htbp]
\begin{center}
\includegraphics[width=0.5\textwidth]{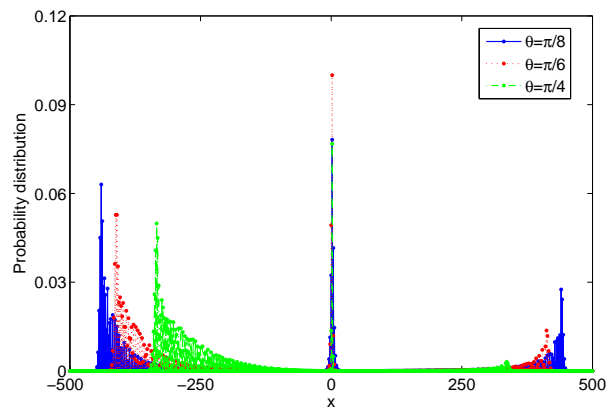}
\end{center}
\caption{\label{fig3} (Color online): The probability distribution of the position in the QRW with the defect residing at $x=2$. Three different values of $\theta$ of the coin operators are chosen. Blue solid, $\theta=\pi/8$, red dotted, $\theta=\pi/6$, green dotted dashed, $\theta=\pi/4$. The initial state of the coin and position is taken as, $|\Phi\rangle_{ini}=(\frac{i}{\sqrt{2}}|0\rangle+\frac{1}{\sqrt{2}}|1\rangle)_c|0\rangle_p$. The phase of the defect, $\phi=1/2$. The time step of the evolution in the QRW is 480.}
\end{figure}

\textit{Even case}. We take the defect occupying the position $x=2$ as an example, the particle starts from the original position ($x=0$). The probability distributions of the position in the QRW with different coin operators are addressed in Fig.~\ref{fig3}.
From the figure, we can find that among the value of the parameters $\theta=\pi/8$, $\pi/6$, and $\pi/4$ of the coin operator, the probability of occupying the position $x=2$ is largest when $\theta$ is taken as $\pi/6$. While, as well known that, when the phase defect emerges at the starting position, the probability distribution of the position in the QRW exhibits a stronger localization behavior around the defect with the increment of $\theta$ in the infinite (even) time~\cite{Andraca_QIP2012, Xue_SRep2014}. However, for the defect resides at the position $x=2$ in our case, the behavior of localization is different.

\begin{figure}[htbp]
\begin{center}
\includegraphics[width=0.5\textwidth]{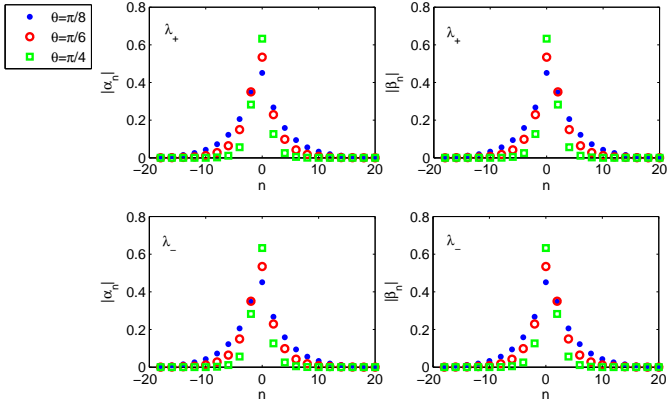}
\end{center}
\caption{\label{fig4} (Color online): The amplitude of the probabilities $\bar{\alpha}_n$ and $\bar{\beta}_n$ with the defect residing at $x=2$. The phase of the defect, $\phi=1/2$. The values of $\bar{\alpha}_n$ and $\bar{\beta}_n$ only emerge at the even number of $n$. Three different values of $\theta$ of the coin operator are considered, blue dotted, $\theta=\pi/8$, red circle, $\theta=\pi/6$, green rectangle, $\theta=\pi/4$. The two figures above describe the amplitudes for the eigenvector $|\psi_+\rangle$, and the two figures below represent the amplitudes for the eigenvector $|\psi_-\rangle$.}
\end{figure}

Now, we begin to elaborate the reason for the localization behavior in our QRW. For each value of these three different $\theta$s ($\theta=\pi/8$, $\pi/6$, and $\pi/4$), we obtain two different eigenvalues ($\lambda_+$ and $\lambda_-$) and the corresponding eigenvectors ($|\psi_+\rangle$ and $|\psi_-\rangle$) by applying Eq.~\ref{evol} and Eq.~\ref{lamb}. The coefficients of $\bar{\alpha}_n$ and $\bar{\beta}_n$ for different eigenvectors are presented in Fig.~\ref{fig4}, the zero point of $x$-axis in our figure represents the position $x=m$ that the defect occupies. The points $\pm2$ of $x$-axis in the figure depict the nearest position with the defect, the points $\pm4$ of the $x$-axis describe the second-nearest position with the defect, and so on. When considering the expressions $\alpha_n$ and $\beta_n$ mentioned in Eq.~\ref{state}, the coefficients $\bar{\alpha}_n$ and $\bar{\beta}_n$ take the values of $\alpha_{n+m}$ and $\beta_{n+m}$, respectively.

The initial state of the coin and position is, $|\Phi\rangle_{ini}=(\cos\varphi\cdot e^{i\delta}|0\rangle+\sin\varphi|1\rangle)_c|0\rangle_p$. The transition probability from the initial state to the localized stationary states of the QRW $|\psi_+\rangle$ and $|\psi_-\rangle$ are, $P(|\Phi\rangle_{ini}\rightarrow|\psi_+\rangle)=|\langle\psi_+|U_\phi^{2N}|\Phi\rangle_{ini}|^2$ and $P(|\Phi\rangle_{ini}\rightarrow|\psi_-\rangle)=|\langle\psi_-|U_\phi^{2N}|\Phi\rangle_{ini}|^2$, respectively. Here, the superscript $2N$ implies that the particle of the QRW takes $N$ evolution operator $U_\phi^2$.
The probability of the particle occupying at position $x=l$ with respect to different localized stationary states $|\psi_\pm\rangle$ is obtained as,
\begin{equation}
\begin{split}
P(n=l)_{|\psi_+\rangle}&=|\lambda_1^N\cdot\bar{\alpha}_{l-m}^{+}(\cos\varphi\cdot e^{i\delta}\bar{\alpha}_{-m}^{+*}+\sin\varphi\cdot\bar{\beta}_{-m}^{+*})|^2\\
                &+|\lambda_1^N\cdot\bar{\beta}_{l-m}^+(\cos\varphi\cdot e^{i\delta}\bar{\alpha}_{-m}^{+*}+\sin\varphi\cdot\bar{\beta}_{-m}^{+*})|^2,\\
P(n=l)_{|\psi_-\rangle}&=|\lambda_2^N\cdot\bar{\alpha}_{l-m}^{-}(\cos\varphi\cdot e^{i\delta}\bar{\alpha}_{-m}^{-*}+\sin\varphi\cdot\bar{\beta}_{-m}^{-*})|^2\\
                &+|\lambda_2^N\cdot\bar{\beta}_{l-m}^-(\cos\varphi\cdot e^{i\delta}\bar{\alpha}_{-m}^{-*}+\sin\varphi\cdot\bar{\beta}_{-m}^{-*})|^2,\label{prob}
\end{split}
\end{equation}
where the defect stays at the position $x=m$. In our depiction of Fig.~\ref{fig3}, the parameters of our initial state $|\Phi\rangle_{ini}$ is taken as, $\varphi=\pi/4$, and $\delta=\pi/2$. When we take the value of the phase $\phi$ of the defect as $1/2$, and the value of $\theta$ is $\pi/8$, $\pi/6$, or $\pi/4$, the localized stationary states $|\psi_+\rangle$ and $|\psi_-\rangle$ both have contributions to the probability $P(n=l)$. Considering that we study the amplitude of the probability at the position of the defect, $l=2$, the associated terms $\bar{\alpha}_n$ and $\bar{\beta}_n$ in Eq.~\ref{prob} are $\bar{\alpha}_0$, $\bar{\alpha}_{-2}$, $\bar{\beta}_0$ and $\bar{\beta}_{-2}$. As addressed in Fig.~\ref{fig4}, although the values $\bar{\alpha}_0$ and $\bar{\beta}_0$ increase with the increase of $\theta$, the values of other amplitudes, $\bar{\alpha}_{-2}$ and $\bar{\beta}_{-2}$, do not go up with the $\theta$ monotonically. The value $\bar{\alpha}_{-2}$ and $\bar{\beta}_{-2}$ with $\theta=\pi/6$ are obviously larger than that with $\theta=\pi/4$, respectively. So in the QRW with one phase defect, the probability of occupying at positon $x=2$ goes up when the parameter $\theta$ of the coin operator changes from $\pi/8$ to $\pi/6$, later it drops down to a smaller value with the parameter $\theta$ changing to $\pi/4$.

\begin{figure}[htbp]
\begin{center}
\includegraphics[width=0.5\textwidth]{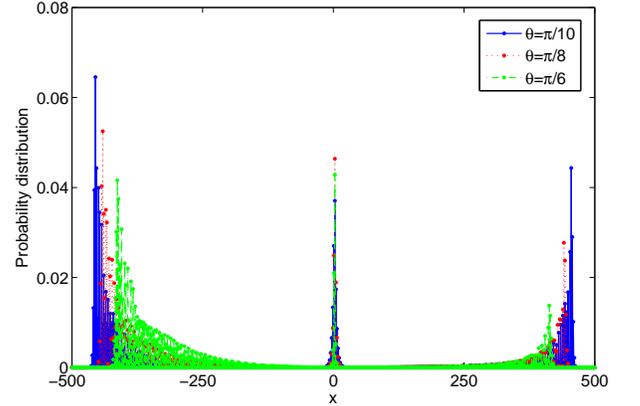}
\end{center}
\caption{\label{fig5} (Color online): The probability distribution of the position in the QRW with the defect residing at $x=3$. Three different values of $\theta$ of the coin operators are chosen. Blue solid, $\theta=\pi/10$, red dotted, $\theta=\pi/8$, green dotted dashed, $\theta=\pi/6$. The initial state of the coin and position is taken as, $|\Phi\rangle_{ini}=(\frac{i}{\sqrt{2}}|0\rangle+\frac{1}{\sqrt{2}}|1\rangle)_c|0\rangle_p$. The phase of the defect, $\phi=1/2$. The time step of the evolution in the QRW is 481.}
\end{figure}

\begin{figure}[htbp]
\begin{center}
\includegraphics[width=0.5\textwidth]{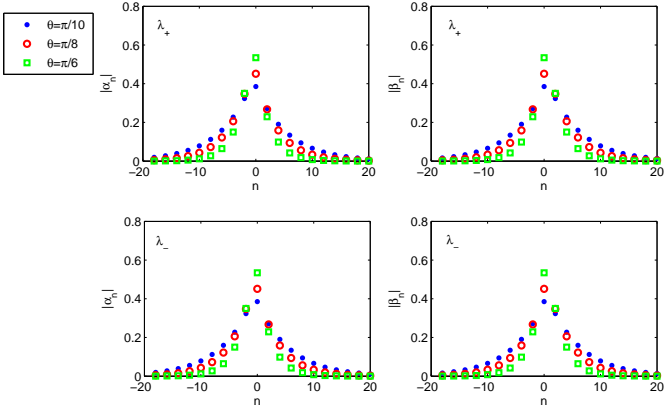}
\end{center}
\caption{\label{fig6} (Color online): The amplitude of the probabilities $\bar{\alpha}_n$ and $\bar{\beta}_n$ with the defect residing at $x=3$. The phase of the defect, $\phi=1/2$. The values of $\bar{\alpha}_n$ and $\bar{\beta}_n$ only emerge at the even number of $n$. Three different values of $\theta$ of the coin operator are considered, blue dotted, $\theta=\pi/10$, red circle, $\theta=\pi/8$, green rectangle, $\theta=\pi/6$. The two figures above describe the amplitudes for the eigenvector $|\psi_+\rangle$, and the two figures below represent the amplitudes for the eigenvector $|\psi_-\rangle$. }
\end{figure}

\textit{Odd case}. We have discussed the effect of coin operators on the probability distribution of the QRW with the defect appearing at the position $x=2$. Next, we will consider the probability distribution of the QRW with the defect occupying at the odd position. We take $x=3$ as the position of the phase defect. The particle starts from the original position, $x=0$. The initial state of the coin and the position is, $|\Phi\rangle_{ini}=(\cos\varphi\cdot e^{i\delta}|0\rangle+\sin\varphi|1\rangle)_c|0\rangle_p$. Three different coin operators are taken with the choices of $\theta$ as, $\theta=\pi/10$, $\pi/8$, and $\pi/6$. The amplitudes of the probabilities for the positions in the QRW are revealed in Fig.~\ref{fig5}.
We can find that the localization of the probability distribution appears at the position $x=3$.
As stated by the previous results, when the position of the single defect is placed at $x=1$, the localization of the probability distribution appears at the position $x=1$, and the amplitude of the probability of occupying this position increases monotonically with the increase of value of $\theta$.~\cite{Kempe_CP2003, Andraca_QIP2012, Xue_SRep2014} But the amplitude of localization at the position $x=3$ does not change larger when the value of $\theta$ increases, see Fig.~\ref{fig5}. The probability at the position $x=3$ with $\theta=\pi/8$ is larger than the probability with $\theta=\pi/6$ and $\theta=\pi/10$. The reason for this phenomena is provided in the following. We start to analyze the amplitudes of the eigenvectors of $U_\phi^2$. Two eigenvalues ($\lambda_{1,2}$) and two localized eigenstates ($|\psi_\pm\rangle$) are obtained. The amplitudes of $\bar{\alpha}_n$ and $\bar{\beta}_n$ for the eigenvectors ($|\psi_+\rangle$ and $|\psi_-\rangle$) are shown explicitly in Fig.~\ref{fig6}.
Considering the probability distribution of the QRW emerges at the odd position in the odd time, we apply one step of evolution $U_\phi$ to the initial state $|\Phi\rangle_{ini}$ and obtain $|\Phi\rangle_{1}$ as,
\begin{equation}
\begin{split}
|\Phi\rangle_{1}=&U_\phi|\Phi\rangle_{ini}\\
=&(e^{i\delta}\cos\varphi\cdot\cos\theta+\sin\varphi\cdot\sin\theta)|0\rangle_c|-1\rangle_p\\&+(e^{i\delta}\cos\varphi\cdot\sin\theta-\sin\varphi\cdot\cos\theta)|1\rangle_c|1\rangle_p.\label{new}
\end{split}
\end{equation}
When taking into account the parameters $\lambda_{\pm}$ and $|\psi_{\pm}\rangle$ are the eigenvalues and eigenvectors of $U_\phi^2$, we obtain the transition probability from the initial state to the localized stationary state $|\psi_+\rangle$ and $|\psi_-\rangle$, $P(|\Phi\rangle_{ini}\rightarrow|\psi_+\rangle)=|\langle\psi_+|U_\phi^{2N+1}|\Phi\rangle_{ini}|^2=|\langle\psi_+|U_\phi^{2N}|\Phi\rangle_1|^2$ and $P(|\Phi\rangle_{ini}\rightarrow|\psi_-\rangle)=|\langle\psi_-|U_\phi^{2N+1}|\Phi\rangle_{ini}|^2=|\langle\psi_-|U_\phi^{2N}|\Phi\rangle_1|^2$, respectively. The superscript $2N+1$ of $U_\phi$ means that the total time step evolution of the QRW is $2N+1$. With the explicit expression
$|\Phi\rangle_{1}$ from Eq.~\ref{new}, the probability at position $x=l$ takes the form as,
\begin{equation}\small
\begin{split}
P(n=l)_{|\psi_+\rangle}&=|\lambda_1^N\bar{\alpha}_{l-m}^{+}\{\bar{\alpha}_{-m-1}^{+*}(e^{i\delta}\cos\varphi\cos\theta+\sin\varphi\sin\theta)\\&+\bar{\beta}_{-m+1}^{+*}
(e^{i\delta}\cos\varphi\sin\theta-\sin\varphi\cos\theta)\}|^2\\
                &+|\lambda_1^N\bar{\beta}_{l-m}^+\{\bar{\alpha}_{-m-1}^{+*}(e^{i\delta}\cos\varphi\cos\theta+\sin\varphi\sin\theta)\\&+\bar{\beta}_{-m+1}^{+*}
(e^{i\delta}\cos\varphi\sin\theta-\sin\varphi\cos\theta)\}|^2,\\
P(n=l)_{|\psi_-\rangle}&=|\lambda_2^N\bar{\alpha}_{l-m}^{-}\{\bar{\alpha}_{-m-1}^{-*}(e^{i\delta}\cos\varphi\cos\theta+\sin\varphi\sin\theta)\\&+\bar{\beta}_{-m+1}^{-*}
(e^{i\delta}\cos\varphi\sin\theta-\sin\varphi\cos\theta)\}|^2\\
                &+|\lambda_2^N\bar{\beta}_{l-m}^-\{\bar{\alpha}_{-m-1}^{-*}(e^{i\delta}\cos\varphi\cos\theta+\sin\varphi\sin\theta)\\&+\bar{\beta}_{-m+1}^{-*}
(e^{i\delta}\cos\varphi\sin\theta-\sin\varphi\cos\theta)\}|^2.
\end{split}
\end{equation}
When the defect resides at position $m=3$, the probability of occupying the position $l=3$ is related to the terms, $\bar{\alpha}_0$, $\bar{\alpha}_{-4}$, $\bar{\beta}_0$, and $\bar{\beta}_{-2}$. Taking into account the parameters addressed in Fig.~\ref{fig5}, we have chosen $\varphi=\pi/4$, and $\delta=\pi/2$. The phase of the defect, $\phi=1/2$. As presented in Fig.~\ref{fig6}, the amplitudes of $\bar{\alpha}_0$ and $\bar{\beta}_0$ is becoming larger with the increase of $\theta$. When considering two other coefficients $\bar{\alpha}_{-4}$ and $\bar{\beta}_{-2}$, these two terms take the largest value for $\theta$ is not $\pi/6$. That is the reason that we can obtain the largest probability at position $x=3$ when the value of $\theta$ is taken as $\pi/8$, given three different values of $\theta$ in Fig.~\ref{fig5}. In contrast, when the defect stays at $x=1$, the probability of locating at position $x=1$ increases with the increase of $\theta$~\cite{Kempe_CP2003, Andraca_QIP2012, Xue_SRep2014}. The localization of the probability distribution in the QRW when the defect occupies the position $x=3$ exhibits different behaviors with that containing the defect at position $x=1$.

Based on the discussion above, it is apparent that the coin operator has a dominant influence on the amplitude of localization of probability distribution in the QRW. By employing the obtained values of coefficients for the localized eigenstates, we can understand why the amplitude of localized probability exhibits the different dependence on the coin operator, when the defect occupies different positions. This non-trivial localized behavior is attributed to the overlap between the initial state and the localized eigenstates associated with the coin operator. Similar analysis on the localization in the QRW can be executed for the defect occupies the position $x\geq4$ also.

So far, we have studied theoretically the localization of probability distribution in the QRW with defects. Such localization of probability distribution can be observed in the experiment as realized in Ref.~\cite{Schreiber_PRL2011, Schreiber_Sci2012}. In their experiments, the Hilbert space for the coin operator is spanned by the polarization degree of the light, and the step evolution is realized with the polarizing beam splitters (PBS) and fiber lines. Different positions in the QRW is revealed with different arriving times of photons in the avalanche photodiodes (APD). By applying the time-dependent signal to the electro-optic modulator (EOM), the phase defect can be introduced into the certain position of the QRW. Considering the QRW with the defect occupying the position $x=2$, we find that the localization of probability distribution is apparent when the particle undergoes 30 steps evolution in the QRW. For the experimental realization mentioned above, the standard QRW with 28 steps evolution has been achieved~\cite{Schreiber_PRL2011}. This experimental realization might provide a platform to observe the localization of the probability distribution in the QRW with defects.

\section{conclusion}\label{IV}

In summary, we have studied the localization of the position distribution in the QRW on an infinite chain. When the single phase defect is introduced into the position of the QRW, the probability at that position where the defect occupies does not tend to zero in the infinite time limit, and the localization of the probability distribution in the QRW emerges. Later we discuss the effect of different coin operators on the localization of the QRW. Taking the defect residing at the position $x=2$ or $x=3$ as examples, we find that the localized probability of the position where the defect occupies does not go up monotonically with the increase of $\theta$. Such non-trivial $\theta$-dependence of localized probability in the QRW is different from that when the defect locates at the position $x=0$ or $x=1$, in which a trivial monotonic increase of localized probability with $\theta$ is revealed~\cite{Kempe_CP2003, Andraca_QIP2012, Xue_SRep2014}. Our new findings of localization in the QRW with defects deepen our insight into the properties of the QRW, and help us to design quantum algorithms and quantum computation in the QRW.

\section*{Acknowledgments}
We acknowledge the financial support from Young Teachers Academic Starting Plan No. 2015CX04046 of Beijing Institute of Technology.

{}


\begin{thebibliography}{}


\bibitem{Motwani_book95} R. Motwani and P. Raghavan, \textit{Randomized Algorithms} (Cambridge University Press, New York, 1995).

\bibitem{Kempe_CP2003} J. Kempe, Contemp. Phys. \textbf{44}, 307 (2003).

\bibitem{Andraca_QIP2012} S. E. Venegas-Andraca, Quantum Inf. Proc. \textbf{9}, 405 (2010).

\bibitem{Farhi_PRA1998} E. Farhi and S. Gutmann, Phys. Rev. A \textbf{58}, 915 (1998).

\bibitem{Aharonov_PRA1993} Y. Aharonov, L. Davidovich, and N. Zagury, Phys. Rev. A \textbf{48}, 1687 (1993).

\bibitem{Childs_QIP2002} A. M. Childs, E. Farhi, and S. Gutmann, Quantum Inf. Proc. \textbf{1}, 35 (2002).

\bibitem{Strauch_PRA2006} F. W. Strauch, Phys. Rev. A \textbf{74}, 030301(R) (2006).

\bibitem{Shenvi_PRA2003} N. Shenvi, J. Kempe, and K. Birgitta Whaley, Phys. Rev. A \textbf{67}, 052307 (2003).

\bibitem{Keating_PRA2007} J. P. Keating, N. Linden, J. C. F. Matthews, and A. Winter, Phys. Rev. A \textbf{76}, 012315 (2007).

\bibitem{Perets_PRL2008} H. B. Perets, Y. Lahini, F. Pozzi, M. Sorel, R. Morandotti, and Y. Silberberg, Phys. Rev. Lett. \textbf{100}, 170506 (2008).

\bibitem{Childs_PRL2009} A. M. Childs, Phys. Rev. Lett. \textbf{102}, 180501 (2009).

\bibitem{Zahringer_PRL2010} F. Z\"{a}hringer, et al. Phys. Rev. Lett. \textbf{104}, 100503 (2010).

\bibitem{Lovett_PRA2010} N. B. Lovett, S. Cooper, M. Everitt, M. Trevers, and V. Kendon, Phys. Rev. A \textbf{81}, 042330 (2010).

\bibitem{Underwood_PRA2010} M. S. Underwood and D. L. Feder, Phys. Rev. A \textbf{82}, 042304 (2010).

\bibitem{Yalcinkaya_JPA2015} \.{I}. Yal\c{c}{\i}nkaya and Z. Gedik, J. Phys. A \textbf{48}, 225302 (2015).

\bibitem{Wojcik_PRL2004} A. W\'{o}jcik, T. {\L}uczak, P. Kurzy\'{n}ski, A. Grudka, and M. Bednarska, Phys. Rev. Lett. \textbf{93}, 180601 (2004).

\bibitem{Ribeiro_PRL2004} P. Ribeiro, P. Milman, and R. Mosseri, Phys. Rev. Lett. \textbf{93}, 190503 (2004).

\bibitem{Inui_PRA2004} N. Inui, Y. Konishi, and N. Konno, Phys. Rev. A \textbf{69}, 052323 (2004).

\bibitem{Anderson_PR1958} P. W. Anderson, Phys. Rev. \textbf{109}, 1492 (1958).

\bibitem{Konno_QIP2010} N. Konno, Quantum Inf. Proc. \textbf{9}, 405 (2010).

\bibitem{Chandrashekar_PRA2011} C. M. Chandrashekar, Phys. Rev. A \textbf{83}, 022320 (2011).


\bibitem{Crespi_Nat2013} Andrea Crespi et al. Nature Photonics, \textbf{7}, 322 (2013).

\bibitem{Xue_NJP2014} P. Xue, H. Qin, B. Tang and B. C Sanders, New J. Phys. \textbf{16}, 053009 (2014).

\bibitem{Nicola_PRA2014} F. De Nicola et al. Phys. Rev. A \textbf{89}, 032322 (2014).

\bibitem{Xue_PRL2015} P. Xue et al. Phys. Rev. Lett. \textbf{114}, 140502 (2015).

\bibitem{Wojcik_PRA2012} A. W\'{o}jcik et al. Phys. Rev. A \textbf{85}, 012329 (2012).

\bibitem{Li_PRA2013} Z. J. Li, J. A. Izaac, and J. B. Wang, Phys. Rev. A \textbf{87}, 012314 (2013).

\bibitem{Lam_PRA2015} H. T. Lam and K. Y. Szeto, Phys. Rev. A \textbf{92}, 012323 (2015).

\bibitem{Zhang_PRA2014} R. Zhang, P. Xue, and J. Twamley, Phys. Rev. A \textbf{89}, 042317 (2014).

\bibitem{Zhang_QIP2014} R. Zhang and P. Xue, Quantum Inf. Proc. \textbf{13}, 1825 (2014).

\bibitem{Li_SRep2015}Z. J. Li and J. B. Wang, Sci. Rep. \textbf{5}, 13585 (2015).

\bibitem{Xue_SRep2014} P. Xue, H. Qin, and B. Tang, Sci. Rep. \textbf{4}, 4825 (2014).

\bibitem{Xue_PRA2015} P. Xue et al, Phys. Rev. A \textbf{92}, 042316 (2015).

\bibitem{Schreiber_PRL2011} A. Schreiber, et al. Phys. Rev. Lett. \textbf{106}, 180403 (2011).

\bibitem{Schreiber_Sci2012} A. Schreiber et al. Science \textbf{336}, 55 (2012)


\end{thebibliography}
\end{document}